\documentclass[preprint,pteplogo]{ptephy_v2}

\preprintnumber{XXXX-XXXX} 
\usepackage{hyperref}

\usepackage{graphics} 
\usepackage{CJK}
\usepackage{comment}
\usepackage[normalem]{ulem}
\usepackage{xcolor}

\begin{document}

\title{First measurement of $\phi$ meson production in 30~GeV proton-nucleus reactions via di-electron decay at J-PARC}



\newcommand\KEK {Present address: IPNS, KEK, Tsukuba 305-0801, Japan}
\newcommand\KEKacc {Present address: Accelerator Laboratory, KEK,  Tsukuba 305-0801, Japan}
\newcommand\KYOTOF {Present address: Institute for the Promotion of Excellence in Higher Education, Kyoto University,  Kyoto 606-8501, Japan}
\newcommand\NIAS {Present address: Institute for Innovative Science and Technology, Nagasaki Institute of Applied Science, Nagasaki, 851-0121, Japan}
\newcommand\ICEPP {Present address: International Center for the Elementary Particle Physics (ICEPP), University of Tokyo, Tokyo, 113-0033, Japan}
\newcommand\RCNP {Present address: Research Center for Nuclear Physics (RCNP), Osaka University, Ibaraki, 567-0047, Japan}
\newcommand\NAGOYA {Present address: Department of Physics, Nagoya University, Nagoya 464-8602, Japan}
\newcommand\TOHOKU {Present address: Department of Physics, Tohoku University, Sendai 980-8578, Japan}
\newcommand\RIKENkaitaku {Present address: RIKEN Cluster for Pioneering Research, RIKEN, Wako 351-0198, Japan}
\newcommand\RIKENnishina {Present address: RIKEN Nishina Center, RIKEN, Wako 351-0198, Japan}
\newcommand\Deceased {Deceased}

\makeatletter
\renewcommand\maketitle{\par
   \begingroup

\renewcommand{\@fnsymbol}[1]{\ifcase##1\or \hbox{*}\or \titdagger\or a\or b\or c\or d\or e\or f\or g\or 
h\or i\or j\or k\or l  \else\@ctrerr\fi}
    
    \thispagestyle{titlepage}%
    \setcounter{footnote}{0}
    \renewcommand\thefootnote{\textsuperscript{\@fnsymbol\c@footnote}}%
    \def\@makefnmark{\rm\@thefnmark}%
    \long\def\@makefntext##1{
      {${\@thefnmark}$}{##1}}%
    \global\@topnum\z@   
    \@maketitle
      \markboth{\@shortauthorlist}{\@shorttitle}
    \@thanks
  \endgroup
  \@afterindentfalse
  \@afterheading
  \setcounter{footnote}{0}%
  \global\let\maketitle\relax
  \global\let\@maketitle\relax}
\makeatother

\author{
Satomi~Nakasuga$^{1,2}$\thanks{E-mail: nakasuga.satomi@gmail.com},
Yuhei~Morino$^3$,
Kazuya~Aoki$^3$,
Yoki~Aramaki$^4$,
Daichi~Arimizu$^1$,
Sakiko~Ashikaga$^5$, 
Wen-Chen~Chang$^6$, 
Ren~Ejima$^7$, 
Hideto~En’yo$^4$, 
Dairon~Rodriguez~Garces$^{8,9}$, 
Johann~M.~Heuser$^{9}$, 
Ryotaro~Honda$^3$, 
Masaya~Ichikawa$^{2,3}$, 
Daichi~Ishii$^1$, 
Shunsuke~Kajikawa$^{10}$, 
Jo~Kakunaga$^{11}$, 
Koki~Kanno$^3$, 
Daisuke~Kawama$^4$,
Yusuke~Komatsu$^{3}$,
Takehito~Kondo$^7$,
Yusuke~Hori$^1$, 
Hikari~Murakami$^{11}$,
Tomoki~Murakami$^4$, 
Ryotaro~Muto$^3$, 
Shunnosuke~Nagafusa$^{1,4}$, 
Wataru~Nakai$^3$, 
Megumi~Naruki$^1$, 
Yuki~Obara$^{11}$,
Shuta~Ochiai$^1$, 
Makoto~Ogura$^1$, 
Kyoichiro~Ozawa$^3$, 
Adrian~Rodriguez~Rodriguez$^9$, 
Hiroyuki~Sako$^2$, 
Susumu~Sato$^2$, 
Michiko~Sekimoto$^3$, 
Kenta~Shigaki$^7$, 
Kazuki~Suzuki$^1$,
Tomonori~Takahashi$^5$, 
Tomohiro~Taniguchi$^1$, 
Maksym~Teklishyn$^9$, 
Alberica~Toia$^{8,9}$, 
Rento~Yamada$^7$, 
Kanako~Yamaguchi$^1$, 
Yorito~Yamaguchi$^7$, 
Shogo~Yanai$^1$, 
and Satoshi~Yokkaichi$^4$

{\bf (J-PARC E16 collaboration)}
}

\affil{
$^1${ Department of Physics, Kyoto University, Kyoto 606-8502, Japan }\\
$^2${ Advanced Science Research Center, Japan Atomic Energy Agency, Tokai 319-1195, Japan }\\
$^3${Institute of Particle and Nuclear Studies (IPNS), High Energy Accelerator Research Organization (KEK), Tsukuba, 305-0801, Japan}\\
$^4${RIKEN Nishina Center for Accelerator-Based Science, RIKEN, Wako, 351-0198, Japan}\\
$^5${Research Center for Nuclear Physics (RCNP), Osaka University, Ibaraki, 567-0047, Japan}\\
$^6${Institute of Physics, Academia Sinica, Taipei, 11529, Taiwan}\\
$^7${Physics Program and International Institute for Sustainability with Knotted Chiral Meta Matter (WPI-SKCM$^2$), Hiroshima University, Higashi-Hiroshima, 739-8526, Japan}\\
$^8${Goethe Universität, Frankfurt am Main, 60438, Germany}\\
$^9${GSI Helmholtzzentrum für Schwerionenforschung GmbH, Darmstadt, 64291, Germany}\\
$^{10}${Department of Physics, Tohoku University, Sendai, 980-8578, Japan}\\
$^{11}${Department of Physics, University of Tokyo, Tokyo, 113-0033, Japan}\\

}





\begin{abstract}
We present the first measurement of the production of the $\phi$ meson in 30~GeV proton-nucleus interactions on carbon and copper targets via the di-electron decay channel. 
The measurement was conducted at the high-momentum beamline of the J-PARC Hadron Experimental Facility, which was commissioned in 2020. 
The $e^+e^-$ pairs were detected using the E16 spectrometer, during a commissioning run of the J-PARC E16 experiment.
The $\phi$ mesons are successfully reconstructed on all experimental targets.
The obtained yields are converted to the total production cross section, assuming a kinematical distribution of the event generator JAM. 
The total cross sections derived are 
2.0 $\pm$ 0.9 (stat.) $\pm$ 1.0 (syst.)~mb on the carbon target and 
10.3 $\pm$ 4.4 (stat.) $\pm$ 4.4 (syst.)~mb on the copper target. 
The mass-number dependence of the cross section is discussed using the parameter $\alpha$, defined as $\sigma \propto A^\alpha$, resulting in $\alpha = $ 0.99 $\pm$ 0.38 (stat.) $\pm$ 0.34 (syst.). 
The extrapolation to $A=1$, which means that the cross section of proton-proton reactions, is in good agreement with the existing measurements at comparable energies. 
\end{abstract}

\subjectindex{D25,D33}

\maketitle

\section{Introduction}

In recent years, intermediate collision energies of around 30 GeV (corresponding to a center of mass energy of $\sqrt{s}=7.7$~GeV) have been actively explored due to the interest in high-density QCD matter. 
In high-energy collisions, perturbative QCD (pQCD) provides a reliable description, where the coupling of the strong interaction decreases as the momentum transfer becomes large.
In general, inclusive hadron production at high energies is well reproduced within the QCD factorization, 
demonstrating the successful hadronization description of the string fragmentation.
On the other hand, in the recently explored intermediate energy region, we need to take into account non-perturbative reaction mechanisms. Since such effects cannot be derived directly from QCD, input from experimental data becomes essential. 
Moreover, since the low-energy applicability of string fragmentation is phenomenologically determined, experimental input is also important for constraining its validity.

In intermediate energy heavy-ion collisions, several signals of enhancement of strange hadron productions have been observed~\cite{phik,kpi}.
Regarding $\phi$-meson production, the NA49 Collaboration observed the unexpectedly broad rapidity distribution~\cite{na49}, and further measurements with different systems have been reported in recent years~\cite{na61pp,na61arsc}. In this context, measurements in proton–nucleus systems may provide complementary information to these results.
In addition, the enhancement of strangeness production has also been discussed in relation to the restoration of chiral symmetry at finite density~\cite{elena1,elena2}. 
From this perspective, measurements in proton–nucleus collisions, reflecting normal nuclear density, may also be informative.

In 2020, the high-momentum beamline was constructed at the J-PARC Hadron Experimental Facility~\cite{hirose}. 
As the first program in this beamline, the J-PARC E16 experiment was launched to study the in-medium spectral modification of the $\phi$ meson, with significantly higher statistics than those achieved in the precedent experiment KEK-PS E325~\cite{muto} via di-electron decay channel. 
In recent years, high-statistics measurements of dilepton spectra have also suggested the possibility of observing signals of degeneracy of chiral partners~\cite{sasaki1,sasaki2,ejima}. 
The E16 setup is also valuable from this perspective. 

In this paper, we report the first measurement of the $\phi$ meson production in 30‑GeV proton-nucleus reactions with copper and carbon targets, using the pilot data of the J-PARC E16 experiment taken in 2024. 
The dependence of the production cross section on the nuclear mass number is commonly expressed using the parameter $\alpha$, defined as follows: 
\begin{equation}\label{eq:alpha}
\begin{split}
\sigma_{pA}(A)  \propto A^\alpha
\end{split}
\end{equation}
to investigate the production mechanism. $\alpha<1$ means suppressed production by existing nuclear matter, in particular, $\alpha=2/3$ corresponds to soft production on the surface of the target nucleus, while $\alpha>1$ indicates enhancement of production. 
Understanding the $\phi$ production mechanism is a fundamental prerequisite 
not only for investigating QCD matter in the intermediate energy,
but also for evaluating the in-medium effects in the J-PARC E16 experiment~\cite{ichikawa}.
However, the mass number (A) dependence measurements of $\phi$ production in proton-nucleus reactions via the dilepton decay channel, which is a penetrating probe of nuclear matter, have been limited: at low energies of the KEK-PS~\cite{tabaru} and at high energies of CERN SPS~\cite{na60}. 
Moreover, there are no data in this energy region with a dilepton decay channel, making the present measurement important. 

The structure of this paper is as follows. Section 2 describes the experimental setup, including the high-momentum beamline, the E16 spectrometer, and the data used in this analysis. Section 3 explains the procedure for spectral reconstruction, the evaluation of efficiencies, and the method of beam normalization. Section 4 presents the results, including the reconstructed spectra and total cross sections obtained under the assumptions of the $\phi$ kinematic distributions derived from an event generator. The A-dependence of the cross section is compared with other experiments, and systematic uncertainties are summarized. Finally, Section 5 provides the conclusions and outlook for future measurements.

\section{Experimental setup}

\subsection{high-momentum beamline}
At the J-PARC Hadron Experimental Facility, a new beamline, called the ``High Momentum Beamline," was constructed to directly supply a small fraction of the 30~GeV primary proton beam to experiments. Operation of this beamline began in May 2020. The beamline is approximately 150~m long and branches off from the existing primary beamline.
A Lambertson magnet is the central component enabling this branching. This magnet is designed with a cross-section that contains both a strong magnetic field region and a field-free hole. The majority of the primary beam (approx. 99.98\%) travels straight through the field-free hole. However, by intentionally directing a small fraction  (approx. 0.02\%) of the beam (the tail) into the magnetic field region, its trajectory is bent, causing it to branch off. This method allows for the simultaneous operation of the primary beamline and the high-momentum beamline.

The beam intensity of the high-momentum beamline is precisely controlled by the vertical position of the primary beam at the Lambertson magnet. A higher incident position causes a larger fraction of the beam to enter the magnetic field, thus increasing the high intensity of the high momentum beamline. The operational beam intensity is $1\times10^{10}$ protons/spill, delivered in 2-second extractions within a 4.24-second cycle. The intensity is monitored by an ion chamber located just upstream of the beam dump. The absolute value was calibrated using an activation method based on $^{24}$Na production in aluminum (Al) and copper (Cu) foils~\cite{beam1,beam2}.

To ensure a constant beam intensity during extraction, a ``ramp control" system was implemented. This system dynamically controls the current of five vertical bending magnets during the spill. This technique allows the beam's incident position on the Lambertson magnet to be set intentionally low at the start of the spill (resulting in low intensity). The position is then gradually raised during the spill (increasing the intensity), which flattens the beam intensity during the spill.

The beam profile at the experimental target is 2~mm wide (horizontal) and 0.5~mm wide (vertical). This profile was determined by moving beam across a 0.3~mm wide, L-shaped Beryllium-Copper (BeCu) rod located near the target and measuring the number of secondary particles yield as a function of the beam position.

\subsection{J-PARC E16 spectrometer}

\begin{figure}[htbp]
  \centering
  \includegraphics[width=\linewidth]{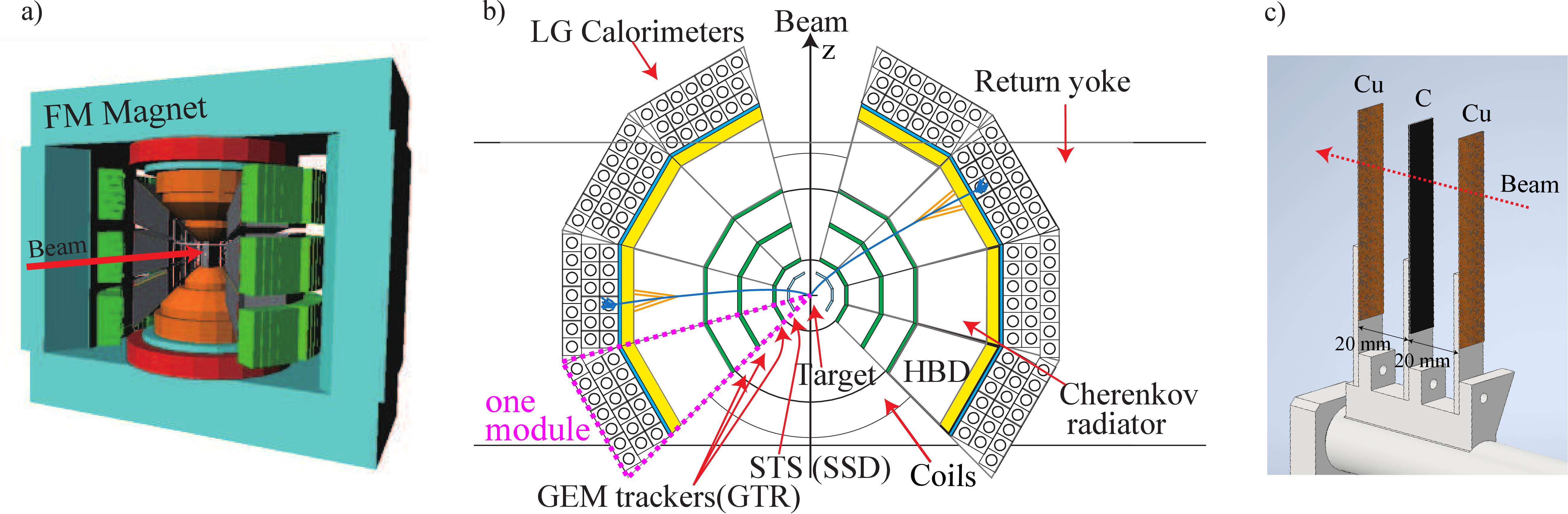}
  \caption{
  a) Three-dimensional schematic view of the proposed E16 spectrometer in the spectrometer magnet (called 'FM magnet').
  b) Cross-sectional view of the E16 spectrometer. The depicted eight detector modules are already constructed in the present data taking.
  c) Enlarged view of the experimental targets.
  }
  \label{fig:det}
\end{figure}

The J-PARC E16 spectrometer was specifically designed to achieve high-rate capability and large geometrical acceptance, enabling the collection of a substantial dataset of vector-meson events using a high-intensity proton beam of $1\times10^{10}$ particles per spill. Figure~\ref{fig:det} illustrates a three-dimensional schematic of the dipole magnet, a cross-sectional view of the E16 spectrometer, and an enlarged depiction of the experimental target region. 
The 30~GeV proton beam traverses three thin target foils: two copper and one carbon. The target configuration is summarized in Table~\ref{tab:tgt}. To minimize the radiation length of each target, two copper foils were employed instead of a single thick foil. Furthermore, the use of multiple thin targets, combined with vertex selection requiring particles to originate from the same target, ensures accurate identification of interaction events even under high-rate conditions.

\begin{table}[t]
\caption{Configuration of experimental targets. The z-axis, as shown in Fig.~\ref{fig:det}(b), is aligned with the beam direction, and the origin is defined at the center of the dipole magnet. }
\label{tab:tgt}
\centering
\begin{tabular}{cp{1.7cm}p{2.7cm}p{2.2cm}p{2.2cm}p{2.2cm}c}
\hline
Target & Position z (mm) & Nominal thickness (mm) & Thickness (mg/cm$^2$) & Interaction length (\%) & Radiation length (\%) & Width (mm)\\
\hline
\hline
Cu & $+20$, $-20$ & 0.08 & 70.8 & 0.052 & 0.55 & 10\\
C & 0 & 0.5 & 89.6 & 0.102 & 0.21 & 10\\
\hline
\end{tabular}
\end{table}

The E16 spectrometer consists of a dipole magnet called the FM magnet, 4-layer tracking devices, and two types of electron identification (EID) detectors. 
Due to the geometry of the FM magnet, the field is uniformly strong near the central pole and decreases towards the outside. 
The strength of the magnetic field is 1.77~T at the center. 
The four types of detectors are grouped to form a unit called a module. One module includes four detectors, covering $\pm$ 15$^{\circ}$ vertically and $\pm$ 12--14$^{\circ}$ horizontally, as shown in magenta dotted lines in Fig.~\ref{fig:det}(b). 
Eight modules are arranged symmetrically on either side of the beam direction, with four modules positioned on the left and four on the right arms. 
For this analysis, only four forward modules are used because most of the produced $\phi$ mesons are detected there. 

The tracking detector is composed of three layers of gas chambers equipped with gas electron multipliers (GEMs)~\cite{sauli}, and single layer of silicon strip detector. The GEM chambers are called the GEM tracker (GTR), whose detailed configuration is found in Ref.~\cite{TMnim}. 
Each GEM chamber has readout strips with two dimensions, x (horizontal) and y (vertical). 
The silicon strip detector and the readout system, called the silicon tracking system (STS)~\cite{aokiproc2025,aokicbm} was developed in collaboration with the GSI group of the CBM experiment at FAIR. 
Two-dimensional readout strips are installed; however, only the horizontal coordinate is utilized in this analysis. Track reconstruction is performed using hit-position information from seven readout planes. 

For electron identification, a gas Cherenkov detector called hadron blind detector (HBD)~\cite{phenixnim} and a lead-glass calorimeter (LG) are developed. The radiator of the HBD is CF$_4$ and the threshold momentum for background pions is 4.2 GeV/$c$. 
The HBD developed by E16~\cite{kannonim} detects 11 photoelectrons on average for electron tracks. The LG has a depth of eight radiation length to properly separate electron signals from the pion signals. The energy resolution is 17\% for electrons at 1 GeV/$c$. 
Details for the configuration and performance of the HBD and LG are referred to Refs.~\cite{nakasuganim1,nakasuganim2}. 

To distinguish the pile-up signals, 
waveform information is recorded for all detectors~\cite{takahashi} except the STS, whose signals are sufficiently segmented in both time and space.

\subsection{Trigger and event selection}\label{sec:dataset}
Data acquisition was carried out from May to June 2024 with a typical beam intensity of $1\times10^{10}$ protons per spill. For this analysis, a total of 2.28$\times 10^{4}$ spills of data were used. The trigger condition was defined using signals from the LG, HBD, and the outermost layer of the GTR~\cite{mrg}. 
An electron candidate was first defined by a triple coincidence of GTR, HBD, and LG trigger hits, requiring coarse position matching among the three detectors. At least two such candidate were required in coincidence. Additionally, the opening angle between two distinct tracks was calculated using the LG hit positions. To suppress electron pairs originating from photon conversions, a wide opening angle—typically greater than $45^{\circ}$—was imposed. 
In addition, trigger veto is applied for the high multiplicity period of the GTR trigger signals
to reduce the fake triggers from pile-ups of multiple collisions. 
This veto period is strongly correlated to the instantaneously high intensity of proton beam~\cite{nagafusaproc}. This effect on the evaluation of the cross section is explained in Sec.~\ref{sec:nproton}.

In the offline analysis, positively charged tracks were selected exclusively in the left arm of the spectrometer, while negatively charged tracks were selected in the right arm to explicitly favor pairs with a wide opening angle. Reconstructed tracks with momenta in the range of 0.5–2.4~GeV/$c$ were used, in accordance with the requirements imposed by the LG performance for electron identification.

\section{Analysis}

\subsection{Event reconstruction}\label{sec:eventreconst}

The mass reconstruction procedure consists of four steps: single-track reconstruction, association with electron-identification (EID) detectors, determination of the interaction target using track pairs, and calculation of the invariant mass. 
First, track candidates are reconstructed from all possible combinations of hit positions in the GTR and STS. The Runge–Kutta fitting method is then applied to candidates that successfully pass a preliminary quadratic fit. Next, these tracks are extrapolated to the detection planes of the HBD and LG, where position matching is required between electron hits in the EID detectors and the projected track positions. For the HBD, signals exceeding six photoelectrons are required to ensure high rejection power. For the LG, the energy-over-momentum ratio~\cite{nakasuganim2} must be greater than 0.53. At this stage, only one track candidate with the smallest chi-square value is retained for each HBD hit to eliminate spurious tracks. 
All remaining tracks are paired by combining positive and negative candidates. To determine the interaction target, the Runge–Kutta method is applied again, requiring the two tracks to share a common vertex. The target closest to the reconstructed vertex along the z-axis is then selected. Finally, the momenta of the two tracks are refined through a fit constrained to the target z-position defined in the previous step (0 or $\pm20$~$\mathrm{mm}$).

\subsection{Efficiency evaluation}\label{sec:eff}

The track reconstruction efficiency was evaluated using low-intensity calibration data and the embedding method. First, the intrinsic detection efficiencies of the GTR and STS were estimated by reconstructing tracks while excluding one layer and using only the remaining three. To ensure high track purity under this condition, calibration data with a beam intensity approximately one-tenth that of the experimental data were employed. The single-track efficiency, which requires all layers to be used, was found to be 77\% on average. 
Next, the reconstruction efficiency was evaluated under realistic background conditions using the embedding method. In this approach, waveform samples from the GTR were embedded into real experimental data, and the same reconstruction procedure described in Section~\ref{sec:eventreconst} was applied. The waveform samples were extracted from low-intensity calibration data. 
The reconstruction efficiency varies with event multiplicity; therefore, all recorded events were scanned to account for multiplicity dependence. The average efficiency for the three targets, evaluated at the most frequent multiplicity, was 31\%. 
The systematic uncertainty, estimated from the multiplicity-dependent correction and the cut for the vertex reconstruction, was found to be 13\% relative to the efficiency.

The association width on the EID detection plane is determined using the residual estimated from the low-intensity calibration data. 
The efficiency of the EID association and the offline timing cut, collectively referred to as the “analysis cut”, was evaluated to be 95\%.

The EID efficiency was evaluated separately for the HBD and LG. For the HBD, trigger-unbiased tracks contained in the experimental dataset were used. The HBD response to electrons selected by the LG was examined, accounting for position-dependent variations. The position-averaged single-electron efficiency was determined to be 55\%. 
The LG efficiency was evaluated primarily using the energy-over-momentum distribution, applying the threshold of E/p $>$ 0.53 as described in Section~\ref{sec:eventreconst}. The resulting efficiency of single-electrons was 96\% on average, taking into account the position- and incident-angle-dependent response of the calorimeter and the effect of charge sharing among channels.

\subsection{The number of protons}\label{sec:nproton}

The number of beam protons was monitored using an ion chamber located upstream of the beam dump, which measures the total charge during each spill with an accuracy of 10\%. As noted in Section~\ref{sec:dataset}, a trigger veto was applied based on the multiplicity of GTR trigger hits.
The veto period is correlated with the observed time-dependent beam intensity~\cite{nagafusaproc}. Therefore, the number of protons irradiated during the veto period was excluded from the total proton count measured by the ion chamber. These remaining protons are referred to as effective protons.
To determine the number of effective protons, the vetoed period was monitored using a streaming TDC~\cite{hul}. Since the ion chamber does not provide the time structure of the beam, LG single-hit counts were used as an alternative. It has been confirmed that the LG single-hit rate is proportional to the beam intensity~\cite{nakasuganim1}. The ratio of LG counts during the non-vetoed period to the total LG counts was calculated, yielding a typical value of 71\%. The number of effective protons was then obtained by multiplying this ratio by the total proton count measured by the ion chamber, spill-by-spill.

When accounting for DAQ live time, the same procedure was applied. The time structure of the trigger request and accept signals was recorded using the streaming TDC, allowing the determination of the time-dependent DAQ live time, defined as the ratio of accepted to requested triggers. The time-dependent product of the number of effective protons and the DAQ live time is referred to as recorded protons.
A total of $1.21\times10^{14}$ recorded protons was obtained, with a relative uncertainty of 2\% arising from the granularity of the time dependence considered
, in addition to a 10\% uncertainty from the ion chamber.
The time-averaged DAQ live time was typically evaluated to be 85\%.

\section{Result and Discussion}
\subsection{Mass spectra}\label{sec:mass}

\begin{figure}[htbp]
  \centering
  \includegraphics[width=1.0\linewidth]{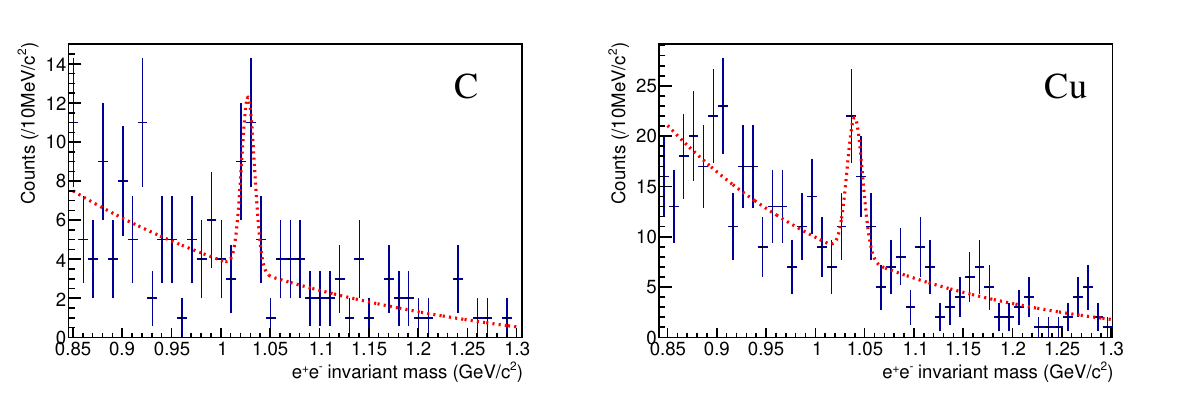}
  \caption{Invariant mass spectra on carbon and copper experimental targets.}
  \label{fig:mass}
\end{figure}

The reconstructed mass spectra for two target nuclei are shown in Fig.~\ref{fig:mass}. These spectra are not corrected for acceptance.
Each spectrum was fitted using a Gaussian function to represent the resonance shape and an exponential function plus a constant term for the background. 
In this fit, six parameters were floated: the amplitude, mean, and sigma of the Gaussian function; the amplitude and slope parameter of the exponential function; and the constant term.
The fitting results are indicated by red dotted lines in Fig.~\ref{fig:mass}.
The background shape was determined from this fit, and the signal yield was extracted as the integral of the mass histogram after subtracting the background contribution. 

To evaluate systematic uncertainties, an alternative background model using a quadratic function was examined, along with a signal shape derived from simulation through the embedding method, as described in Section~\ref{sec:eff}.
The same integral region for counting the number of signals is used in all fittings.
In addition, several different sets of center positions of the bins are examined to evaluate the systematic uncertainties. 
Under these conditions, the resulting significance varied in the range of 2.0--4.5 for the carbon target and 3.2--4.9 for the copper target.
We finally assign the systematic uncertainty as the largest difference from the representative value.
We also confirmed that contamination from other targets or materials is negligibly small from the obtained vertex distribution.

The evaluated yields are summarized in Table~\ref{tab:yield}. 
We also show the fitting chi-square and the significance, where it is calculated as $S/\sqrt{B}$, with $S$ denoting the number of signal events (yield) and $B$ the number of background events.

\begin{table}[t]
\caption{Obtained yields on carbon and copper experimental targets.}
\label{tab:yield}
\centering
\begin{tabular}{ccccc}
\hline
Target & Yield & Significance & $\chi^{2}/ndf$\\
\hline
\hline
C  & 11.9 $\pm$ 5.6 $\pm$ 5.2 & 2.9 & 40.5/39\\
Cu & 23.6 $\pm$ 10.2 $\pm$ 8.5 & 3.4 & 37.8/39\\
\hline
\end{tabular}
\end{table}

The uncertainties of the absolute mass scale is up to 2\%.
This deviation can be reasonably attributed to the accuracy in the geometrical alignments among different detector modules and to the uncertainty in the vertex determination.
The uncertainty in the vertex determination affects the efficiency during the Runge–Kutta fitting of the pair tracks. This systematic uncertainty is estimated to contribute 5\% and is included as part of the ``Track Reconstruction".

\subsection{Systematic uncertainties}

The systematic uncertainties for the evaluation of the total cross section and the parameter $\alpha$ are summarized in Tables~\ref{tab:syst}. 
For $\alpha$, only the target-dependent systematic uncertainties arising from the target thickness and yield, namely the first two rows in Table~\ref{tab:syst}, are considered.

The beam intensity and DAQ efficiency with trigger veto are described in Section~\ref{sec:nproton}, and the track reconstruction, analysis cut, and electron identification are discussed in Section~\ref{sec:eff}.
The yields are given in Section~\ref{sec:mass}. 
The trigger efficiency includes the timing match of the coincidence of the trigger signals and the trigger reconstruction efficiency at the electronics.
The acceptance takes into account the uncertainty due to the position dependence of the detection efficiency.

\begin{table}[t]
\caption{Summary of systematic uncertainties for the total cross sections.}
\label{tab:syst}
\centering
\begin{tabular}{ccc}
\hline
 & C (\%) & Cu (\%)\\
 \hline
 \hline
Yield & 44 & 36\\
Target thickness & 0.1 & 0.1\\
Beam intensity & 10 & (same as left)\\
DAQ efficiency with trigger veto & 2 & (same as left)\\
Branching ratio & 1.1~\cite{pdg} & (same as left)\\
Trigger efficiency & 1 & (same as left)\\
Track reconstruction & 13 & (same as left)\\
Analysis cut & 6 & (same as left)\\
Electron identification & 6 & (same as left)\\
Acceptance & 13 & (same as left)\\
\hline
Total & 50 & 43\\
\hline
\end{tabular}
\end{table}

\subsection{Total cross section}

To examine the total production cross section for $\phi$-meson, we adopt a kinematic distribution generated by the event generator JAM (v1.011)~\cite{jam} for the evaluation of detector acceptance. The momentum distribution of $\phi$ at the decay point in the code was extracted. 
A $\phi$ meson decays isotropically into an electron and a positron taking 
account of the internal radiative corrections (including internal bremsstrahlung, vertex correction, and vacuum polarization)~\cite{irc}. 
The experimental effects on the electron and positron in the materials of the detectors and experimental targets, including the Coulomb multiple scattering, the Bethe-Bloch type energy loss and bremsstrahlung, were evaluated by the detector simulation using the Geant4 toolkit (v4.9.5p01)~\cite{geant1,geant2}. 
From these considerations, the probability for electrons and positrons to enter the detector acceptance was obtained.

\begin{figure}[htbp]
  \centering
  \includegraphics[width=0.8\linewidth]{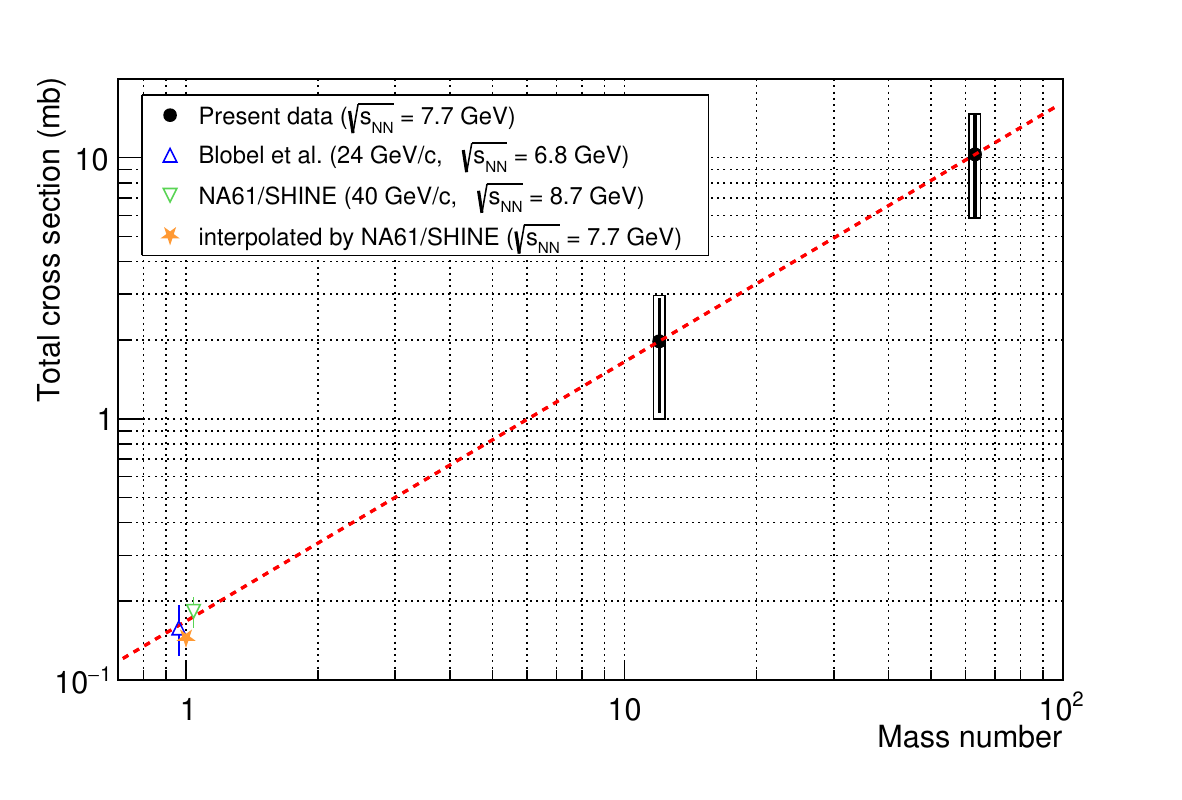}
  \caption{
  Mass number dependence of the estimated total cross section.
  The filled black circles represent the results of the present measurement. The black bars indicate the statistical uncertainties, while the unfilled black boxes denote the systematic uncertainties of the total cross section.
  The linear extrapolation of these two data points is shown as a red dashed line.
  The results from measurements with the other energies are shown as a triangle and an inverted triangle.
  The orange star is explained in the text.
  }
  \label{fig:adep}
\end{figure}

Consequently, the measured yields were converted into total cross sections, which are determined to be 
2.0 $\pm$ 0.9 (stat.) $\pm$ 1.0 (syst.)~mb for the carbon target and 
10.3 $\pm$ 4.4 (stat.) $\pm$ 4.4 (syst.)~mb for the copper target.
For copper, the acceptances of the two targets (at $z = -20$ and $+20$~mm) were calculated separately and then summed.
From the ratio of these cross sections, the extracted $\alpha$ is 
0.99 $\pm$ 0.38 (stat.) $\pm$ 0.34 (syst.). 
Figure~\ref{fig:adep} presents the estimated cross sections as a function of mass number. The red dashed line indicates a linear extrapolation of these two points to lower mass numbers. Although no data exist for 30~GeV proton-proton reactions, measurements at beam momenta of 24~GeV/$c$~\cite{blovel} and 40~GeV/$c$~\cite{na61pp} are available. The 24~GeV/$c$ data are shown as a blue triangle and the 40~GeV/$c$ data as a green inverted triangle, plotted with a slight horizontal offset from unity for clarity.

The NA61/SHINE experiment measured the $\phi$ meson production cross sections in pp reactions at 40~GeV/$c$, 80~GeV/$c$, and 158~GeV/$c$, and compared their energy dependence with other world data in the center-of-mass energy range from $\sqrt{s_{NN}}$ = 6.8~GeV to 52.5~GeV~\cite{na61pp}.
As a result, they suggest the linear dependence of the center-of-mass energy $\sqrt{s_{NN}}$.
Assuming this dependence,
the interpolated cross section at 30~GeV is shown in Fig~\ref{fig:adep} as an orange star.
The $\alpha$ parameter derived from the present measurement, although with large uncertainties, shows good agreement with the cross sections obtained from existing proton-proton measurements.

Finally, the $\alpha$ parameter obtained in the present measurement is compared with results at other beam energies using di-lepton decay channels. The values for each beam energy are summarized in Table~\ref{tab:ene}. Previous measurements include the 12~GeV data from KEK-PS~\cite{tabaru} and the 400~GeV data from the CERN SPS NA60 experiment~\cite{na60}. The present result shows an $\alpha$ value consistent with those at other energies. This suggests that, at a beam energy of 30~GeV, there is no compelling evidence to exclude $\phi$-meson production scaling with the mass number ($\alpha \sim 1$).

\begin{table}[t]
\caption{Comparison of $\alpha$ parameter with different production energies.}
\label{tab:ene}
\centering
\begin{tabular}{ccc}
\hline
Beam energy (GeV) & $\alpha$\\
\hline
\hline
12~\cite{tabaru} & 0.937 $\pm$ 0.049 $\pm$ 0.018\\
30 & 0.99 $\pm$ 0.38 $\pm$ 0.34\\
400~\cite{na60} & 0.906 $\pm$ 0.011 $\pm$ 0.025\\
\hline
\end{tabular}
\end{table}

\section{Summary and Outlook}

We report the first measurement of $\phi$ meson production in 30~GeV proton-nucleus reactions on carbon and copper targets via the di-electron decay channel. 
This is the first physics result obtained using the newly commissioned high-momentum beam line at J-PARC. 
The invariant-mass spectra of $\phi$ mesons were successfully reconstructed on each target, and the corresponding yields were extracted from these spectra. Using kinematic distributions generated by the event generator JAM, the yields were converted into total cross sections. 
The $\alpha$ parameter, derived from the ratio of the two cross sections, exhibits good agreement with results from proton-proton reaction experiments.
Moreover, the obtained $\alpha$ value is consistent with previous measurements at different beam energies, indicating no significant deviation from a production mechanism proportional to the nuclear mass number ($\alpha \sim 1$).
In addition, this measurement serves as an important milestone for future high-statistics study of in-medium vector meson properties at J-PARC.

\section*{Acknowledgment}

The authors express their gratitude to the members of the J-PARC accelerator and the Hadron Experimental Facility. In particular, we thank the Hadron Beamline Group for their efforts in constructing and operating the J-PARC high-momentum beamline. 
We appreciate the staff of ELPH, Tohoku University, RIKEN RIBF, and KEK PF-AR test beamline for their support in the test experiments of detectors. 
We are grateful to the staff of the KEK Electronics System Group for their help in the development and testing of the readout circuits, and also
staff of the Advanced Manufacturing Support Team in RIKEN RAP, the Mechanical Engineering Center in KEK, and the Engineering Service Department in JAEA
for their help in the detector construction.
We acknowledge the efforts of the staff at KEKCC, RIKEN-CCJ, and RIKEN HOKUSAI (RB23009) for their support with our data analysis and data archiving.
A part of the authors were supported by the RIKEN SPDR and JRA programs, and Grant-in-Aid for JSPS Fellows JP18J20494, JP23KJ1223, JP23KJ1650. This project was supported by  MEXT/JSPS KAKENHI Grant Numbers JP19654036, JP19340075, JP21105004, JP26247048, JP15H05449, JP15K17669, JP18H05235, JP20H01935, JP20H05647, JP21H01102, JP23H05440, JP24H00236, and National Science and Technology Council of the Republic of China (Taiwan).

\let\doi\relax


\end{document}